\documentclass{article}
\usepackage{graphicx}

\begin{document}

\setcounter{page}{0}
\begin{titlepage}
\title{Spin Glass and ferromagnetism in disordered Kondo lattice}
\author{S. G. Magalh\~aes$^a$, F. M. Zimmer$^a$, P. R. Krebs$^b$\\[3mm]         
{\it $^a$Departamento de Fisica - UFSM}\\         
{\it 97105-900 Santa Maria, RS, Brazil}\\
{\it $^b$Instituto de F\'{\i}sica e Matem\'atica -
 UFPEL}\\ {\it 96010-900, Pelotas, RS,Brazil}}
\date{}
\maketitle
\thispagestyle{empty}
\begin{abstract}
\noindent 
The competition among spin glass (SG), ferromagnetism and Kondo effect has been 
analysed in a Kondo lattice model where the inter-site coupling $J_{ij}$ between the localized magnetic 
moments is given by a generalized Mattis model \cite{Mattis} which represents 
an interpolation between ferromagnetism and a highly disordered spin glass. 
Functional integral techniques with of Grassmann fields has been used to obtain the partition function.
The static approximation and the replica symmetric ansatz has also been used. 
The solution of the problem is presented as a phase diagram temperature $T$ {\it versus} $J_K$ (the 
strength of the intra-site interaction). If $J_K$ is small, 
for decreasing temperature there is a second order transition from a paramagnetic to 
a spin glass phase
For lower temperatures, a first order transition 
appears where solutions for the spin glass order parameter and the local magnetizations 
are simultaneously non zero. For very low temperatures, the local magnetizations becomes thermodinamically stables. 
For high $J_K$, the Kondo state is dominating. These results could be helpful 
to clarify the experimental situation of $CeNi_{1-x}Cu_{x}$ \cite{Gomez-Sal1,Gomez-Sal2}. 
\end{abstract}
\vspace*{0.5cm}
\end{titlepage}

The effects of the disorder in the competition between RKKY interaction 
and the screening of  localized moments due to the Kondo effect are quite novel. 
In particular, when the disorder and the RKKY interaction are combined to produce  
frustration. For instance, the alloy $Ce$$Ni_{1-x}$$Cu_{x}$ 
has been investigated by bulk methods \cite{Gomez-Sal1,Gomez-Sal2} showing that 
for low  $Ni$ content there is an antiferromagnetic phase. For high $Ni$ content, 
the Kondo effect is dominating, it leads to a reduction of the magnetic moments. However, 
in the intermediated region, when the temperature is decreased, it appears a 
spin glass-like (SG) phase at $T_{f}$. The SG-like region becomes larger when $Ni$ doping increases.  
 There is the onset of the ferromagnetism (FE) at $T_{c}$, below the spin glass-like region 
($T_{c} < T_{f}$).

Recently, a theoretical effort has been done to understand the emergence of a SG  
and ferromagnetic phases in a Kondo lattice model \cite{Alba1,Magal1} with 
a random Gaussian inter-site coupling among the localized spins. The partition function 
has been obtained within the path 
integral formalism by writing the spins operators as bilinear combinations of Grassmann
fields.
The inter-site coupling is treated using the Sherrington-Kirkpatrick (SK) approach. The obtained phase diagram shows the Kondo state, FE and 
SG solutions. Nevertheless, the FE solution appears always above the SG one in temperature 
($T_{c} > T_{f}$). That is 
a clear indication that the high degree of frustration of the SK approach seems to be not 
adequate to tackle the experimental situation in the $Ce$$Ni_{1-x}$$Cu_{x}$. 

We suggest that a approach where it is possible to interpolate from weak to 
strong frustration would be more adequate to address the experimental features  of the $Ce$$Ni_{1-x}$$Cu_{x}$ alloy. 
That can be achieved by using the coupling between spins given by the generalization of 
the Mattis model \cite{Mattis}:
\begin{eqnarray}
J_{ij}=\frac{J}{N}\sum_{\mu=1}^{p}\xi_{i}^{\mu}\xi_{j}^{\mu}
\label{e4}~,
\end{eqnarray}    
where $\xi^{\mu}_{i}=\pm 1$ ($i=1...N$, $\mu=1,...,p$) are random independent variables which follows 
the distribution
\begin{eqnarray}
P(\xi^{\mu}_{j})=1/2\delta_{\xi^{\mu}_{j},+1}+1/2\delta_{\xi^{\mu}_{j},-1}
\label{e5}~.
\end{eqnarray}
It is known that when $p=1$ for classical spins with no magnetic field, it 
is recovered the limit of trivial disorder of the Mattis model 
with no frustration \cite{Mattis}. 
When $p=N$, in the thermodynamic limit ($N\rightarrow \infty$), one has a strong 
frustration  similar to the S-K approach \cite{Amit1}. However, a method originally introduced 
to deal with complex systems \cite{Amit1}, allows one to access the region between this two limits.    
In that approach, it is chosen configurations $\{{\xi^{\mu}_{i}}\}$ which 
minimizes the free energy. From that choice, it is possible to reconstruct the 
couplings $J_{ij}$ to study the phase transitions present in the problem.      
In this method, $p=O(N)$ and the ratio $a=p/N$ is finite 
in the 
thermodynamic limit.  
Thus, $a$ is a parameter  controlling the 
degree of frustration. 

Therefore, the purpose of this work is to combine both methods 
from Refs. \cite{Alba1,Amit1}. The static approximation 
and the replica symmetry ansatz has 
allowed to find the free energy at mean field level  
in terms of SG order parameter $q$, the Kondo order parameter $\lambda$,  
the local susceptibility $\overline{\chi}$ \cite{Alba1} and  
$m^{\mu}=\frac{1}{N}\sum_{i}\xi^{\mu}_{i}<S_{i}>$.  
For $\mu=1$, this parameter indicates the presence of Mattis states which is 
thermodynamically equivalent to ferromagnetism \cite{Amit1}.    

The model is the Kondo lattice 
\cite{Alba1} with  the random intersite coupling $J_{ij}$ given as Eq. (\ref{e4}): 
\begin{eqnarray}
H=\sum_{k,\sigma}\epsilon_{k} n^{c}_{k\sigma}+ \sum_{i,\sigma}\epsilon_{0} n^{f}_{i\sigma}+J_{K}\sum_{i}[S^{+}_{f,i}s^{-}_{i}\nonumber\\
+S^{-}_{f,i}s^{+}_{i}]+\sum_{i,j}J_{ij}S^{z}_{fi}S^{z}_{fj}
\label{e2}
\end{eqnarray}
\begin{figure}     
\centering     
\includegraphics[angle=270,width=12.cm]{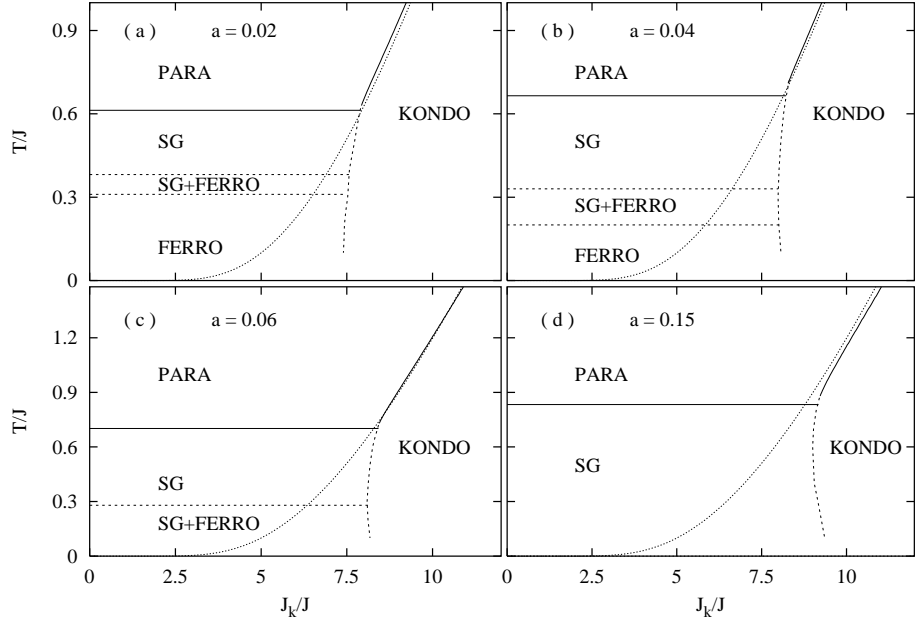}     
\caption{ The phase diagram $T/J$ {\it versus} $J_{K}/J$ for several 
values of $a$ showing the phases SG (spin glass), FERRO (ferromagnetism) and KONDO 
(the Kondo state). In the region SG+FERRO there is the coexistence between 
SG and FERRO. The first order transition is represented by the dashed line. The
dotted means the "pure" Kondo temperature \cite{Alba1}.}      
\label{fig1} 
\end{figure}  

The free energy is found as (the details will be provided elsewhere \cite{Magal3}) :
\begin{eqnarray} 
\beta f=2\beta J_{K}\lambda^{2}
+\frac{a\beta\bar{\chi}}{2}[\frac{1+\beta(q-\bar{\chi})}{1-\beta\bar{\chi}}]
+\frac{\beta}{2}(m^{1})^{2}\nonumber\\
+\frac{a}{2}\ln[1-\beta\bar{\chi}]-
\int^{+\infty}_{-\infty} Dz \ln\{ \int^{+\infty}_{-\infty} Dw \exp\{ 
\frac{1}{\beta D}
\int^{+\beta D}_{-\beta D} dx
\nonumber\\
\langle\langle \ln[2(\cosh\frac{x+h}{2}+
\cosh\sqrt{\Delta^{2} +(\beta J_{K}\lambda)^{2}} )]
\rangle\rangle_{\xi}\}
\label{e33}
\end{eqnarray}
where $<<...>>_{\xi}$ (the average over the $\xi$'s) can be found with Eq.
(\ref{e5}), $Dz=dz\exp{(-z^2/2)}/\sqrt{2\pi}$, $\Delta\equiv\Delta(x,h)=(x-h)/2$ and  
\begin{eqnarray}
h=\sqrt{\beta\frac{2-\beta\bar{\chi}}{1-\beta\bar{\chi}}}w+
\sqrt{ \frac{\beta^2aq}{1-\beta\bar{\chi}}}z+\beta m^{1}\xi
\label{e35}~.
\end{eqnarray} 
In the Eq. (\ref{e33}), it has been used a constant density of states for the conduction electrons,
$\rho(\epsilon)=\frac{1}{2D}$ for $-D<\epsilon < D$.

The Fig. (1) shows the result of the numerical analysis for the order parameters in a phase 
diagram temperature $T/J$ {\it versus} $J_{K}/J$ for several values of $a$.
In the Fig. (1.a) ($a=0.02$), for high temperature and small $J_{K}$, one gets  paramagnetism. 
When the temperature is decreased, there is a second order transition to a SG phase. At low 
temperature appears a first order transition to the FE order with a coexistence region 
indicate as SG+FE in the Fig (1). For high $J_{K}$, appears a new order corresponding to the Kondo state. 
When the parameter $a$ is increased (see Figs. (1.b) and (1.c)), the SG stability 
region is also increased. Finally, for $a=0.15$ (see Fig. (1.d)),  the FE solution disappears 
which recovers the results from Ref. (\cite{Alba1}). Therefore, if 
one associates the $Ni$ content with both $J_{K}$ and the degree of frustration 
$a$, the sequence of phases in Fig.1 has the correct order in temperature as the 
experimental results obtained by bulk methods \cite{Gomez-Sal2}.  
That could also explain the increase of the SG-like region with the increase of $Ni$ 
observed in the experimental diagram.            

%
%
%
%

%
%
%
%
\end{document}